\documentclass[aps,prl,twocolumn,superscriptaddress,groupedaddress]{revtex4}
\usepackage{subfig}
\usepackage{graphicx}  
\usepackage{dcolumn}   
\usepackage{bm}        
\usepackage{amssymb}   
\usepackage{slashed}
\usepackage{graphicx}				
\usepackage{amsmath}
\usepackage{mathtools}
\usepackage{tikz,pgf}
\usepackage{comment}
\usepackage{asymptote}
\usetikzlibrary{arrows,backgrounds}
\usetikzlibrary{fit,scopes,calc,matrix,positioning,decorations.pathmorphing}
\usepackage[all]{xy}
\usepackage{yfonts}

\begin{document}
\title{Strings and missing wormhole entanglement}
\author{Andrei T. Patrascu}
\address{ELI-NP, Horia Hulubei National Institute for R\&D in Physics and Nuclear Engineering, 30 Reactorului St, Bucharest-Magurele, 077125, Romania}
\begin{abstract}
I show that holographic calculations of entanglement entropy in the context of AdS bulk space modified by wormhole geometries provide the expected entanglement magnitude. This arises in the context of string theory by means of additional geometric structure that is seen by the string in its bulk evolution. The process can be described as a net entanglement flow towards stringy geometry. I make use of the fact that as opposed to quantum field theory, strings have additional winding mode states around small extra dimensions which modify the area computation given by the standard application of the Ryu-Takayanagi entanglement entropy formula.
\end{abstract}
\maketitle
Wormholes are very special solutions of Einstein's field equations that persist in the modern interpretations of quantum gravity (string theory) [1]. Heuristically speaking, to keep a wormhole open at least some violations of the averaged weak energy condition must take place [2]. There are several quantum field theoretical phenomena that are capable of this [3], [4]. When we wish to compute entanglement entropy the tool of choice in the context of holography is the Ryu-Takayanagi formula [5], [6]. Its prescription recommends us to calculate the entanglement entropy associated to a boundary region as the area of the minimal surface reaching inside the AdS bulk space in such a way that the surface in the bulk is homotopically equivalent to the boundary surface and it is anchored on it. This prescription is known to offer a formula which suffers a series of corrections both from higher order geometric or quantum phenomena [7] and from string theoretical effects [8]. While it is not particularly surprising that string geometry will play a role in entanglement entropy estimates, when dealing with wormhole geometries the standard holographic evaluation of entanglement entropy is lower than what would be expected [9]. While there exist cases where wormhole geometries are suitably manipulated to provide the entanglement entropy we might expect, such manipulations are often artificial and come with other fundamental issues [10]. 
In this article I argue for the thesis that entanglement is a fundamental property of string theory as well, a property that affects in a fundamentally different way the physics of wormholes and hopefully other phenomenological predictions that may be easier to test experimentally. 
Indeed, entanglement plays an important role in condensed matter systems where topologically non-trivial structures are used to classify certain large scale properties like topological insulators, topological superconductors or materials with long range entanglement. Calculating the accurate entanglement entropy therefore is a critical aspect in understanding the large scale behaviour of topological quantum materials. As entanglement entropy is a measure of quantum entanglement and wormholes are topologically non-trivial solutions of Einstein's field equations, understanding the entanglement properties of wormholes may become important in understanding large scale properties of quantum materials with non-trivial topological structure. Condensed matter analogues to wormhole geometries have been studied in graphene structures [14] leading to the construction of graphene wormholes in which short nanotubes act as a connection between two graphene sheets. It has been shown in [14] that such graphene wormholes can be described by an effective theory of two Dirac fermion fields in the corresponding curved wormhole geometry. A proper estimation of the entanglement obtained between the two ends could therefore establish new macroscopic properties of graphene monolayers with topological structure [15]. Understanding how string phenomena in the bulk affect the calculation of entanglement entropy on the boundary may be the first step towards better understanding non-perturbative string theoretical effects in a holographic context. Moreover, such an observation may reveal new quantum states of matter where long range entanglement is stronger than previously thought, provided one can associate the geometry of the quantum system to a wormhole geometry i.e. a topologically non-trivial object. 

In this sense, the main departure of string theory from the well known quantum field theoretical analysis is the fact that strings are extended objects which have non-trivial winding modes. String theory also requires the presence of additional compact dimensions. Therefore, a string in its dynamics, will perceive spacetime geometry  rather differently, compared to a point-like particle. Also, the calculation of areas as required by the Ryu-Takayanagi formula will be different when the involved objects are strings. Heuristically speaking, given the fact that the wormhole throat can be of the size of the Planck scale [3], [4] it is not unexpected that a measuring device analysing the area of a geometry which contains wormholes will be based on strings and stringy geometry. Analysing string theory even in the case of simple toroidal compactification shows that for certain areas, the phenomenon of gauge symmetry extension leads to a transition from the field theoretical gauge group $U(1)\times U(1)$ to an extended group $SU(1)\times SU(1)$ as shown in [11]. This is a first indication that strings do not see the same geometry as basic quantum field theory objects would. Indeed, a string moving on a periodically compactified dimension may have winding number $w=0$ case in which it will not wrap around the extra dimension. However, by processes involving joining or splitting of strings, such a state can decay into states with winding number $w=\pm 1$ case in which a topologically non-trivial winding state of the string will appear. Introducing the supplemental area arising from such higher winding states both explains the extended gauge symmetry and induces additional entanglement entropy, not visible from any quantum field theoretical perspective. Of course the state $w=\pm 1$ is entangled in a way that cannot be seen from a point-like particle point of view. To express the Ryu-Takayanagi formula, consider a $d+1$ dimensional asymptotically anti-de-Sitter spacetime $\mathcal{M}$. Its conformal boundary is given by $\partial\mathcal{M}$. Consider a spacelike foliation of the conformal boundary along an assumed timelike Killing field. One may according to the Ryu-Takayanagi prescription divide this foliation into regions $\mathcal{A}$ and $\mathcal{B}$. The boundary between these regions will be denoted $\partial\mathcal{A}$. For a quantum field theory on $\partial\mathcal{M}$ the entanglement entropy associated to, say, region $\mathcal{A}$ suffers from ultraviolet divergencies as QFT has an infinite number of degrees of freedom. The leading divergence of the entanglement entropy will scale as the area of the boundary $\partial\mathcal{A}$
\begin{equation}
S_{\mathcal{A}}=\alpha\frac{A(\partial\mathcal{A})}{\epsilon^{d-2}}
\end{equation}
By extending this construction into the bulk we obtain a minimal surface from where we can extract by means of purely geometrical arguments the associated entanglement entropy 
\begin{equation}
S_{\mathcal{A}}=\frac{A(S_{min})}{4G_{N}^{(d+1)}}
\end{equation}
However, it has been shown in [7] that this formula is corrected by terms involving further entanglement of the bulk region delimited by the minimal area and the rest of the bulk. Quantum corrections will also play a role and several divergencies will have to be properly renormalised by means of suitable counter-terms. The resulting formula for the quantum corrections to the entanglement entropy is 
\begin{equation}
\begin{array}{c}
S_{c}=S+S_{q}\\
\\
S_{q}=S_{bulk-ent}+\frac{\delta A}{4 G_{N}}+<\Delta S_{Wald}>+S_{conter}\\
\end{array}
\end{equation}
where the first term is the bulk entanglement, the second is the change in the area due to a quantum shift of the classical background, the third term is of the type given by the Wald entropy formula, and finally the last term is the counter-term component. 
This formula has been constructed in the context of quantum field theories and the contributions to the entanglement entropy are specific therefore to such theories. If we consider a region $A$ and compute the entanglement entropy between this region and the other parts of the system, if there is a gravity dual to the field theory we can compute the entanglement entropy by determining the minimal area surface that starts on the region $A$ of the boundary of the bulk and goes inside the bulk. This result is the first order approximation in the expansion in terms of $G_{N}$. This comes from the classical physics in the bulk. Quantum corrections from the bulk can also be considered and they have been found to be generated by the bulk entanglement entropy. The minimal surface that enters the bulk and is having $A$ as the boundary limit divides the bulk into two regions: the region that is connected to the boundary $A_{b}$ and the region that is not, which is its bulk complement. The quantum corrections from the bulk are given by the entanglement entropy between $A_{b}$ and the rest of the bulk. If this is the approximation we wish to take then the theory in the bulk can be regarded also as an effective field theory living on a fixed background geometry for which we can compute the entanglement entropy of the region $A_{b}$ as if we were working with a quantum field theory. 
On another side, there are other ways of generalising the formula for entanglement entropy, by means of a more accurate usage of the replica trick. In particular, the idea of entropy, which has been extended from its original thermodynamic meaning to an informational meaning based on the density matrix, has been used in the work of Gibbons and Hawking in order to obtain the area formula for gravitational entropy. In particular, a thermodynamical interpretation of the Euclidean gravity solutions with an $U(1)$ isometry has been obtained. If one considers Euclidean solutions with given boundary conditions one obtains the solutions invariant under an U(1) symmetry. In this way we can regard these solutions as the result of the computation of the partition function of a quantum theory in a classical approximation. Classically, the boundary can be chosen as any surface on which we put boundary conditions. In particular, this can be extended to solutions without a U(1) symmetry and the entropy resulting for a density matrix representative for such a case would be computed by means of the replica trick. In the case in which we do have U(1) symmetry and have a Gibbons-Hawking type computation then the minimal surface will have zero extrinsic curvature. 
These aspects have however been discussed in ref. [19], [20], [21], [22]. 
The definition of entanglement entropy in a string context becomes difficult for several reasons. Due to the absence of local observables and due to the non-factorisation of Hilbert spaces in quantum field theory, the idea of defining the entanglement entropy by taking a partial trace on one side is complicated. The usual divergence of entanglement entropy is usually related to the existence of strong correlations across small distances and therefore a barrier (a hypersurface) that would separate two regions or allow us to split the Hilbert space would involve strong correlations across it that would have to, eventually, become finite in a theory of quantum gravity. Of course, the non-existence of local observables in quantum gravity (and in fact in any theory involving a realistic gravity) is a difficulty when discussing entanglement entropy by usual methods. We can instead focus on the algebra of local observables, and there it would be possible to form such a separation, however, causality would imply that the two algebras, located in causally separated regions, would completely commute. But because of Shur's lemma, such an algebra could not provide an irreducible representation on a Hilbert space unless the Hilbert space factorises again, hence we go back from where we started. Moreover, if we were to define the problem on a Rindler space and would try to associate one algebra of observables to the right Rindler wedge and the other to the left, we would get into the situation in which one of the algebras is a von Neumann type III algebra and not a type I as would be the case in a purely factorised system. The divergence of such an entanglement entropy is therefore a property of the algebra and would produce the same divergence no matter what the quantum states are that are involved. However, all this discussion of algebras local observables does not really apply to quantum gravity due to the problem related to local observables and gravity. 
The same problem appears with the definition of a modular Hamiltonian, generating time evolution, in quantum gravity. 
One of the approaches to solving the divergences of entanglement entropy in a quantum field theoretical context was to produce a renormalisation of Newton's constant in an Einstein-Hilbert formulation. However, there is no reason that after such a renormalisation, finite variations in Newton's constant to amount to the same finite variations to the entanglement entropy and therefore there are other contributions that one has to consider for entanglement entropy. 

In the context of probing traversable wormholes by means of strings several situations arise. The equivalent structure here amounts to two conformal field theories with the interaction between them being mediated through the bulk connection. In that case, a string can account for phase transitions in the following way. When the two ends of the string belong to the same patch and hence to the same conformal field theory, the system to which that situation corresponds is a quark-anti-quark pair charged under the same field. A string on the other side extending through the wormhole describes a pair of coloured particles charged under different gauge fields, associated to the two different patches. When however a quark anti-quark pair is considered in each boundary, the system undergoes a phase transition. For small separation each pair of charges exhibits Coulomb interaction, while for large separation the charges in the different field theories tend to pair up [16], [17]. The analysis of string probes in the context of thin shell wormholes and the resulting phase transitions determined there have been discussed in [18].

In this article a different contribution is presented, namely resulting from the idea that a string type object is present in the bulk. Therefore there will be string type corrections in the bulk that will determine entanglement entropy contributions that were not directly visible before. From that point of view, such contributions will appear as corrections on the bulk quantum field theoretical entanglement entropy, which however will affect the higher order corrections obtained by means of the use of the replica trick for a non U(1) symmetric case. 
\begin{figure}
  \includegraphics[width=100pt]{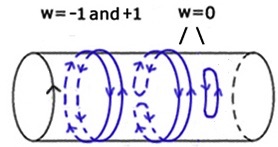}
  \caption{The process of splitting or joining of strings in different winding states and the associated entanglement creation. While the $w=0$ state is homotopical to a point and does not span additional area, the $w=\pm1$ state spans additional area over the extra-dimension which has to be considered in the calculation of entanglement entropy}
  \label{fig:fig1}
\end{figure}
Now let us consider a wormhole throat within the AdS bulk space. Modifying the $AdS_{5}\times S^{5}$ solution to incorporate a wormhole we obtain [9]
\begin{widetext}
\begin{equation}
\begin{array}{c}
ds^{2}=(\frac{H(u)}{cos(u)})^{1/2}(\frac{a^{2}du^{2}}{16 cos^{2}(u)}+a^{2}d\Omega_{5}^{2})+H(u)^{-1/2}(cos(v) (-dt^{2}+dz^{2})+2sin(v)\cdot dt\cdot dz+dx_{1}^{2}+dx_{2}^{2})\\
\\
F_{5}=dt\wedge dz \wedge dx_{1}\wedge dx_{2}\wedge d(H^{-1})+*(dt\wedge dz \wedge dx_{1}\wedge dx_{2}\wedge d(H^{-1}))
\\
\end{array}
\end{equation}
\end{widetext}
where I used the same definitions as in [9]
\begin{widetext}
\begin{equation}
\begin{array}{ccc}
H(u)=(\frac{l}{a})^{4}\cdot (\frac{\pi}{2}-u), & v=\sqrt{\frac{5}{2}}(\frac{\pi}{2}-u), & u=2\cdot arcsin(\frac{r(r^{2}+2a^{2})^{1/2}}{\sqrt{2}(r^{2}+a^{2})})\\
\end{array}
\end{equation}
\end{widetext}
The parameter $l$ describes the $AdS$ radius and the parameter $a$ describes the size of the wormhole. 
The solution above can be reduced to five dimensions with the gravity action and the metric being 
\begin{widetext}
\begin{equation}
\begin{array}{c}
S_{5}=\int dx^{5}\sqrt{-g}(R-\frac{1}{2}\partial_{\mu}\phi \partial^{\mu}\phi+\frac{4}{l^{2}}(5e^{16\alpha \phi/5}-2e^{8\alpha\phi}))\\
\\
ds_{5}^{2}=(\frac{a}{l})^{\frac{10}{3}}(\frac{H(u)}{cos(u)})^{4/3}\frac{a^{2}du^{2}}{16 cos^{2}(u)}+\frac{H(u)^{1/3}}{(cos(u))^{5/6}}(cos(v) (-dt^{2}+dz^{2})+2sin(v)\cdot dt\cdot dz+dx_{1}^{2}+dx_{2}^{2})\\
\\
e^{-6\alpha\phi/5}=\frac{a^{2}}{l^{2}}\sqrt{\frac{H}{cos(u)}}\\
\end{array}
\end{equation}
\end{widetext}

There exists a non-geodesic time-like trajectory which connects the two asymptotic regions [9], [12]. It can be noted that the holographic stress energy tensor for this geometry has as only non-trivial component 
\begin{equation}
T_{tz}=\frac{\sqrt{10}a^{4}}{8\pi\l^{5}}
\end{equation}
which breaks the standard energy condition as the energy is vanishing while the momentum is non-zero [9]. Still, this solution can be embedded in string theory. The 3-dimensional minimal surfaces of the metric given above have been analysed in [9] and an entanglement entropy deficit has been systematically obtained. The holographic entanglement entropy has been calculated for a subsystem $A$ defined by 
\begin{equation}
A=\{(x_{1}, x_{2}, z)| -\infty < x_{1}, z < \infty, -L/2< x_{2} < L/2\}
\end{equation}
The area functional becomes 
\begin{equation}
S=(\frac{a}{l})^{5}\int dx_{1}dz du\frac{H^{1/2}}{cos^{5/4}(u)}\sqrt{L_{1}}
\end{equation}
with 
\begin{equation}
L_{1}=cos(v)\cdot (\frac{a^{2}}{16}\frac{H}{cos^{5/2}(u)}+(x'_{2})^{2})-(t')^{2}
\end{equation}
It is important to notice the size of the wormhole $a$. This will play the role of the radius of the winding dimension seen by the string. As quantum field theoretical arguments show that this size is expected to be of the order of magnitude of the Planck scale [12] (although exceptions have been considered [13]) string dynamics, and particularly string splitting and joining and hence changes in the winding state of the string become possible. Such geometric processes carry away a part of the entanglement entropy we would expect to normally see towards string geometry and hence the calculation must take such string geometry effects into account. Indeed, the effective common area will grow corresponding to the string coupling governing such processes. This observation can be readily written considering the winding of a closed string around a compact direction 
\begin{equation}
\begin{array}{cc}
X(\sigma +2\pi)=X(\sigma)+2\pi a w, & w\in\mathbb{Z}
\end{array}
\end{equation}
with the winding number $w$. The Laurent expansions are given by 
\begin{widetext}
\begin{equation}
\begin{array}{cc}
\partial X(z)=-i(\frac{\alpha'}{2})^{1/2}\sum\limits_{m=-\infty}^{\infty}\frac{\alpha_{m}}{z^{m+1}}, & \bar{\partial}X(\bar{z})=-i(\frac{\alpha'}{2})^{1/2}\sum\limits_{m=-\infty}^{\infty}\frac{\tilde{\alpha}_{m}}{\bar{z}^{m+1}}
\end{array}
\end{equation}
\end{widetext}
The overall change in the coordinate $X$ due to the string moving around the compact dimension is 
\begin{equation}
2\pi a w=\int(dz\partial X + d\bar{z}\bar{\partial}X)=2\pi(\alpha'/2)^{1/2}(\alpha_{0}+\tilde{\alpha}_{0})
\end{equation}
The partition function is therefore interpreted as a sum over all genus one world-sheets in a periodic spacetime, with every non-trivial closed curve capable of wrapping around the compact dimension. This leads to the path integral breaking up into distinct sectors of different topologies labelled by the winding number $w$ and by $m$. The partition function then becomes 
\begin{equation}
2\pi a Z_{X}(\tau)\sum\limits_{m,w=-\infty}^{\infty}exp(-\frac{\pi a^{2}|m-w\tau|^{2}}{\alpha'\tau_{2}})
\end{equation}
with $Z_{X}(\tau)=(4\pi^{2}\alpha'\tau_{2})^{-1/2}|\eta(\tau)|^{-2}$, $\eta(\tau)$ being the Dedekind eta function. The Gaussian path integration is done by writing $X$ as the sum of the classical solution of the known periodicity, plus the quantum piece with periodic boundary conditions [11]. The path integral over the quantum part remains like in the non-compact case, while the compact component to the classical action appears as a correction of the form given in the exponent above
\begin{equation}
S_{compact}=-\frac{\pi a^{2}|m-w\tau |^{2}}{\alpha'\tau_{2}}
\end{equation}
A more intuitive way of looking at this is to think that from the perspective of string theory, the standard $U(1)\times U(1)$ symmetry seen by normal quantum field theories cannot be maintained, as there are certain radii $a=\alpha'^{1/2}$ which correspond to gauge symmetry enhancement. The calculations are fairly standard and can be verified in any string theory textbook [11]. From this perspective, additional winding of the strings leads to more complex types of coupling to scalar fields and hence gives a justification of the fact that geometry must be extended according to string theory in order to deal with entanglement entropy in holographic contexts. One can easily see that a closed string that does not wrap the extra dimension is homotopical to a point and hence does not contribute to the area at larger scales, while the wrapping strings resulting from the splitting ($w=\pm1$) will span an area on the extra-dimension that must be added to the entanglement entropy. It is particularly suggestive to see that string state transitions leading to superpositions of opposite winding modes propagating on compact dimensions are contributing to entanglement entropy in a way that can be represented geometrically in the sense of Ryu-Takayanagi but requires an extension of the bulk area to take into account effects due to opposite winding number strings propagating inside the bulk. These effects become particularly important at compact dimension radii (or equivalently wormhole radii) comparable to the radii where extended symmetries emerge. This phenomenon can be interpreted as a flow of entanglement towards additional dimensions. Considering the ER-EPR duality it could be interesting to speculate whether in the EPR interpretation of wormholes, such extra-dimensional flows of entanglement could have a measurable impact on tabletop entanglement experiments and how such an impact may be interpreted. 

In this article I showed that the standard Ryu-Takayanagi formula for the calculation of entanglement entropy must be extended in order to capture additional structure arising from strings winding around extra-dimensions inside the bulk. Such a phenomenon is the first effect to account for string theoretical geometry inside the bulk and hence may provide a new way of looking at non-perturbative string theory in a holographic context. Moreover, the calculation of entanglement entropy for topologically non-trivial structures may have an impact on the way we understand condensed matter systems where long range entanglement and topologically non-trivial structures are important. 
Much is left for a future publication, like an in depth study of the effect this analysis has on the phase transitions and the preservation of modular invariance. However, as such, this paper brings a new insight into the problem of missing entanglement and the ability of strings to probe wormholes, determine phase transitions, and explain additional entanglement in this context. 


\begin{thebibliography}{99}
\bibitem{1}D. N. Vollick, Class. Quantum Grav. 16, pag. 1599 (1999)
\bibitem{2}M. Bouhmadi-Lopez et al. JCAP11, 007(2014)
\bibitem{3}L. H. Ford, T. A. Roman, Phys. Rev. D 53, 5496 (1996)
\bibitem{4}B. E. Taylor, W. A. Hiscock, P. R. Anderson, Phys. Rev. D 55, 6116 (1997)
\bibitem{5}S. Ryu, T. Takayanagi, Phys. Rev. Lett. 96, 181602 (2006)
\bibitem{6}V. E. Hubeny, M. Rangamani, T. Takayanagi, JHEP 07, 062 (2007)
\bibitem{7}T. Faulkner, A. Lewkowycz, J. Maldacena, JHEP 11, 074 (2013)
\bibitem{8}S. He, T. Numasawa, T. Takayanagi, K. Watanabe, JHEP 05, 106 (2015)
\bibitem{9}M. Fujita, Y. Hatsuda, T. Takayanagi, JHEP 06, 141 (2011)
\bibitem{10}A. Bergman, H. Lu, J. Mei, C. N. Pope, Nucl. Phys. B 810, pag. 300 (2009)
\bibitem{11}J. Polchinski, String Theory, Cambridge University Press, Cambridge (1998)
\bibitem{12}J. Preskill, Nucl. Phys. B 323, pag. 141 (1989)
\bibitem{13}S. Coleman, Nucl. Phys. B 341, pag. 101 (1990)
\bibitem{14}J. Gonzalez, J. Herrero, Nucl. Phys. B, Vol. 825, 3 (2010)
\bibitem{15} Y. Zhang, Y.-W. Tan, H. L. Stormer, P. Kim, Nature 438, pag. 201 (2005)
\bibitem{16} H. Liu, K. Rajagopal, U. A. Wiedemann, Phys. Rev. Lett 98, 182301 (2007)
\bibitem{17} M. Chernicoff, J.  A. Garcia, A. Guijosa, JHEP 0609, 068 (2006)
\bibitem{18} M. Chernicoff, E. Garcia, G. Gilbert, JHEP 019, 10 (2020)
\bibitem{19} A. Lewkowycz, J. Maldacenva, JHEP 090, 08 (2013)
\bibitem{20} R. Das Sumit, A. Kaushal, G. Mandal, K. K. Nanda, M. H. Radwan, S. P. Trivedi, JHEP 0141, 04 (2023)
\bibitem{21} S. N. Solodukhin, Living Rev. Rel. 14, 8 (2011)
\bibitem{22} J. Maldacena, L. Maoz, 	JHEP 0402 (2004) 053
\end{thebibliography}
\end{document}